\documentclass[aps,prd,longbibliography,nofootinbib,amsthm,amsmath,amssymb,amsfonts,notitlepage]{revtex4-1}
\usepackage[utf8]{inputenc}
\usepackage[T1]{fontenc}
\usepackage[english]{babel}
\usepackage{graphicx}
\usepackage{float}
\usepackage{cleveref}
\usepackage{xcolor}
\usepackage{tikz}
\usetikzlibrary{arrows}

\date{\today}

\begin{document}

\title{Decay of the vortex muon}
\author{Pengcheng Zhao}
\email{zhaopch5@mail2.sysu.edu.cn}
\author{Igor P. Ivanov}
\email{ivanov@mail.sysu.edu.cn}
\author{Pengming Zhang}
\email{zhangpm5@mail.sysu.edu.cn}
\affiliation{School of Physics and Astronomy, Sun Yat-sen University, 519082 Zhuhai, China}

\begin{abstract}
Muon decay is self-analyzing: the spectral-angular distribution of the emitted electron reveals the spin orientation of the polarized muon.
Here, we show that the same feature applies to muons in non-plane-wave states and helps reveal the rich polarization opportunities available.
We focus on the so-called vortex states, in which the muon carries a non-zero orbital angular momentum with respect to the average propagation direction
and exhibits a cone structure in the momentum distribution.
We compute the spectrum and the angular distribution of the electrons emitted in decays of vortex muons and show that the most revealing
observable is not the angular distribution but the fixed-angle electron spectra.
Even for very small cone opening angles of the vortex muons, it will be easy to observe significant modifications of the electron spectra
which would allow one to distinguish vortex muons from approximately plane wave muons, as well as to differentiate among various polarization states.
These features will be the key to tracking the evolution of vortex muons in external magnetic fields.
\end{abstract}

\maketitle

\section{Introduction}

Particle physics with carefully engineered non-plane-wave states of initial particles is an emergent field
whose full potential is still to be explored.
Although examples of quantum electrodynamics processes in which the non-plane-wave nature
of colliding beams played important role were known since 1980's \cite{Blinov:1982vp,Kotkin:1992bj,Kotkin:2003jz},
a renewed interest was triggered recently by the experimental demonstration
of the so-called vortex, or twisted, states of photons and, especially, electrons.
A vortex state refers to a monochromatic wave with a helicoidal, corkscrew-like, wavefront
arising from the azimuthal phase factor $\exp(i\ell \varphi_r)$.
Such a wave propagates, on average, along axis $z$ and, due to its swirling current density,
also carries a non-zero orbital angular momentum (OAM) $z$-projection, $\ell\hbar$ per quantum.

Vortex photons are known since decades \cite{Allen:1992zz,Molina-Terriza:2007,Harris:2015,Paggett:2017,Knyazev:2019}
and have become a basis of numerous applications \cite{Zhan:2009,Torres:2011}.
A decade ago, following the suggestion of \cite{Bliokh:2007ec}, vortex electrons were experimentally demonstrated
\cite{Uchida:2010,Verbeeck:2010,McMorran:2011}.
They are now routinely used to probe magnetic properties of matter at the atomic scale,
to excite plasmons, and to test behavior of twisted electrons in external magnetic fields,
see reviews \cite{Bliokh:2017uvr,Lloyd:2017}.
In the past few years, neutral particles such as neutrons \cite{Clark:2015,Sarenac:2018,Sarenac:2019}
and, very recently, atoms \cite{Luski:2021} were also put in vortex states,
opening new promising venues for fundamental physics and applications.

From the high-energy physics perspective, a particle prepared in a Bessel vortex state
carries definite energy, definite longitudinal momentum, definite modulus of the transverse momentum,
and a non-zero OAM.
This OAM represents a new degree of freedom, which can be imposed on the initial state particles
to get additional information about their structure and interactions.
In the past decade, several groups theoretically analyzed high-energy collision processes involving
one or two particles in vortex states \cite{Jentschura:2010ap,Jentschura:2011ih,Ivanov:2011kk,Ivanov:2012na,Karlovets:2012eu,Hayrapetyan:2014faa,Stock:2015yha,Serbo:2015kia,Ivanov:2016oue,Karlovets:2016jrd,Zaytsev:2017lnr,Karlovets:2017qda,Afanasev:2017jdf,Sherwin:2018dah,Afanasev:2019rlo,Ivanov:2019pdt,Ivanov:2019vxe,Afanasev:2020nur,Sherwin:2020vyq,Afanasev:2021uth,Afanasev:2021fda}.
Many non-trivial effects were predicted in these and other publications,
many of which are impossible to achieve in conventional (approximately) plane-wave collisions.
For an overview of such processes with vortex electrons, see review \cite{Bliokh:2017uvr} and references therein.
Also, going beyond vortex states, Refs.~\cite{Karlovets:2016dva,Karlovets:2016jrd} address collisions of particles in general non-plane-wave states, 
showing yet more ways to probe particle structure and dynamics.

In this paper, we add yet another process to this list: the decay of vortex muon.
Although muons in vortex states have not yet been demonstrated, we believe it can be done in the near future.
Anticipating these experiments, we find it timely to theoretically explore the emerging physics opportunities.

When vortex muons are produced, one may want to explore their behavior in external electric and magnetic fields,
in particular in storage rings.
One may wonder whether and for how long the vortex state of the muon survives in external fields,
how the vortex parameters and the spin evolve,
and how this evolution is affected by the inhomogeneities of the magnetic field of the ring.
All these effects can be monitored through the decay of muons $\mu \to e\bar \nu_e \nu_\mu$,
which is a self-analyzing process \cite{Miller:2007kk}:
namely, the correlations between the spectrum and the angular distribution of the emitted electron reveal the orientation of the muon spin.
It is this self-analyzing property which lies at the heart of several generations of
experiments measuring the muon anomalous magnetic moment $g-2$ with increasing precision \cite{Miller:2007kk,Logashenko:2018pcl}.

In this paper, we study how the spectrum and the angular distribution of the electron
reveal the parameters of the {\em vortex} muon, including its polarization state.
We will show that the most sensitive observable is not the angular distribution but the electron spectrum at a fixed emission angle.
Even for very small vortex cone opening angles, this spectrum shows dramatic, easily observable features
which would allow one to distinguish the vortex state from an (approximately) plane wave muon
as well as differentiate among various polarization states of the vortex muon.

The paper is organized as follows. In the next section we briefly review the well known properties
of the plane wave muon decays.
We rewrite the classical results in a form convenient for vortex muon calculations.
In section~\ref{section-vortex-decay}, we remind the reader of how polarized vortex fermions are described,
derive analytical results for unpolarized vortex muon decays and study them with illustrative examples.
In the same section we analyze the decay of a vortex muon in various polarization states and show how they can be detected.
Finally, in section~\ref{section-discussion} we discuss the results, briefly mention experimental prospects,
and draw conclusions.

\section{Plane wave muons}\label{PW}

Muon decay is a standard process worked out in many textbooks \cite{Okun:1982, Donoghue:1992dd}.
When calculating tthe muon decay, we limit ourselves to the Standard Model and neglect the electron mass.
The four-momentum of the initial plane wave muon is denoted as $p^\mu$, its energy and mass are labeled simply as $E$ and $m$.
We will also use the unit vector $\vec n$ in the direction of the muon momentum, so that
$p^\mu = E(1,\, \beta \vec n)$, with $\beta$ being the speed of the muon in units of $c$ and $\gamma = 1/\sqrt{1-\beta^2}$ is its relativistic factor.
The final electron carries the four-momentum $k^\mu = E_e (1, \vec n_e)$. The two neutrinos with four-momenta $q_1^\mu$
and $q_2^\mu$ escape detection but their invariant mass $q^2 = (q_1 + q_2)^2$
affects the electron energy: $q^2 = m^2 - 2 (pk)$. In particular, the maximal electron energy
is attained when the two neutrinos are collinear, so that $q^2=0$ and $2(pk)_{\rm max} = m^2$.

A (pure) polarization state of a plane wave muon is described by a four-vector $s^\mu$
orthogonal to the muon four-momentum $(sp) = 0$ and normalized as $s^2 = -1$.
In the muon rest frame, $s^\mu = (0, \vec s)$, with $\vec s^2 = 1$, where $\vec s$ can be interpreted as the spin
of the muon. In the laboratory reference frame, in which the muon is moving, it can be written as
\begin{equation}
s^\mu = \bigl(\gamma\beta (\vec s\vec n), \ \vec s + (\gamma-1)\vec n(\vec s\vec n)\bigr)\,.
\end{equation}
The Standard Model calculation of the polarized muon decay width summed over the final state polarizations gives
\begin{equation}
d\Gamma_{PW} = \frac{G_F^2}{48\pi^4 E} \frac{d^3 k}{E_e} \cdot \left[(pk)(3m^2 - 4(pk)) - m(sk)(m^2 - 4(pk))\right]\,.
\label{PWresult}
\end{equation}
In the muon rest frame we have $E = m$, $\beta=0$, $\gamma=1$, $(pk) = mE_e$,
and the maximal electron energy $E_{e\rm max} = m/2$ independent of the electron emission angle.
In this frame, denoting $E_e = \epsilon E_{e\rm max}$, we switch to integration over $\epsilon$:
\begin{equation}
d\Gamma_{PW} = \Gamma_0 \cdot \frac{d\Omega}{4\pi} \cdot 2 \epsilon^2 d\epsilon [3-2\epsilon + (1-2\epsilon)\cos\theta_{s,e}]\,,
\quad \Gamma_0 = \frac{G_F^2 m^5}{192\pi^3}\,, \quad \cos\theta_{s,e} = \vec s\,\vec n_e\,.
\label{PWresult-2}
\end{equation}
Integrating over $\epsilon$ from 0 to 1, we obtain the angular distribution $1-(\cos\theta_{s,e})/3$ which reveals the forward-backward asymmetry
with respect to the spin orientation. Finally, integrating over all angles gives the total decay width $\Gamma_0$.

\begin{figure}[H]
	\centering
	\includegraphics[width=0.48\textwidth]{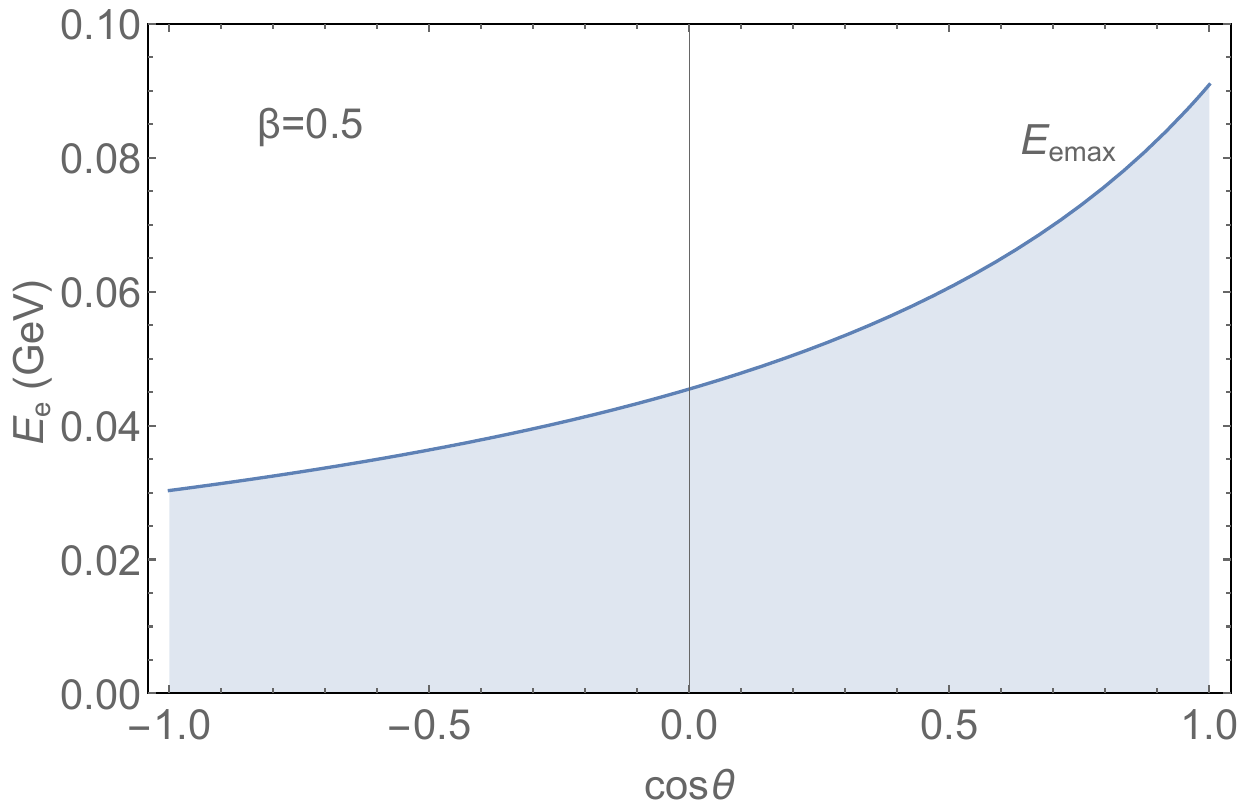}
	\caption{The shaded region shows the range of energies, in GeV, of the electron emitted at polar angle $\theta$ in the decay
		of a muon moving with $\beta = 0.5$. The solid line shows the maximal electron energy as a function of $\theta$.}\label{Fig-EmaxPW}
\end{figure}

In the laboratory frame where the muon is moving, one can express the spectral-angular distribution as
\begin{eqnarray}
d\Gamma_{PW} &=& \frac{G_F^2}{48\pi^4 E}\, E_e dE_e d\Omega\, \Bigl\{
3 m^2 E E_e(1-\beta\, \vec n\vec n_e) - 4 E^2 E_e^2 (1-\beta\, \vec n\vec n_e)^2 - m(sk) [m^2 - 4 E E_e (1-\beta\, \vec n\vec n_e)]\Bigr\}\,,
\label{PWresult-3}\\[1mm]
&&\mbox{where}\quad (sk) = E_e\left[\gamma\beta\, \vec s \vec n - \vec s \vec n_e - (\gamma-1)(\vec s\vec n)(\vec n \vec n_e)\right]\,.
\end{eqnarray}
In this frame, the maximal energy of the electron depends on the emission angle:
\begin{equation}
E_{e\,\rm max} = \frac{m^2}{2E (1-\beta\, \vec n\vec n_e)}\,,\label{Emax}
\end{equation}
which is illustrated in Fig.~\ref{Fig-EmaxPW}.
The electron spectrum extends up to the absolute maximum $E_{e\,\rm abs\, max} = m^2/(2E(1-\beta))$,
and, near this maximal value, only directions close to the muon direction contribute to the decay width.

If we are interested in the angular distribution integrated over all energies,
we can again define $E_e = \epsilon E_{e\,\rm max}$, integrate over $\epsilon$ from 0 to 1 for any given direction,
and present the angular distribution as
\begin{equation}
d\Gamma_{PW} = \Gamma_0\, \frac{d\Omega}{4\pi} \, \frac{1}{\gamma^3}\, \frac{1}{(1-\beta\, \vec n\vec n_e)^2}
\left[1 + \frac{\gamma\beta\, \vec s \vec n - \vec s \vec n_e - (\gamma-1)(\vec s\vec n)(\vec n \vec n_e)}{3\gamma (1-\beta\, \vec n\vec n_e)}\right]\,.
\label{PWresult-4}
\end{equation}
Integration over all angles yields the total decay width $\Gamma = \Gamma_0/\gamma$, which reflects the relativistic time dilatation.

Notice that, anticipating transition to vortex muons, we do not use angles defined with respect to a fixed coordinate frame
but express the angular distribution in terms of the three unit vectors $\vec n$, $\vec n_e$, and $\vec s$.
However for future reference we give the angular distribution of the muon moving alone axis $z$
and polarized along the same axis, $\vec s = \vec n$:
\begin{equation}
d\Gamma_{PW} = \Gamma_0\, \frac{d\Omega}{4\pi} \, \frac{1}{\gamma^3}\, \frac{1}{(1-\beta\, \cos\theta)^2}
\left[1 + \frac{\beta-\cos\theta}{3(1-\beta\, \cos\theta)}\right]\,.
\label{PWresult-4.1}
\end{equation}	
In this expression, $\theta$ is the electron polar angle.

\section{Vortex muon decay}\label{section-vortex-decay}

\subsection{Describing a polarized vortex muon}\label{subsection-describing}

A vortex state of any particle can be constructed as a superposition of plane waves
organized in such a way that a straight line of phase singularity, identified with axis $z$, appears in the coordinate space.
There exist different prescriptions for constructing such a state.
From the computational point of view, the simplest option is given by the so-called Bessel vortex state.
This is a monochromatic solution of the corresponding wave equation written as a superposition of plane waves
with the same energy $E$, same longitudinal momentum $p_z$, same modulus of the transverse momentum $|\vec p_\perp| = \varkappa$
but with different azimuthal angles $\varphi_p$, see further definitions and normalization conditions
in \cite{Jentschura:2010ap, Jentschura:2011ih, Ivanov:2011kk, Karlovets:2012eu}.
As we run through all the plane wave components inside a Bessel state, 
the vectors of their momenta revolve around axis $z$ and cover
the surface of a cone with the opening angle $\theta_0$, defined as
\begin{equation}
\cos\theta_0 = p_z/|\vec p|\,, \quad \sin\theta_0 = \varkappa/|\vec p|\,.\label{cone}
\end{equation}
In realistic situations, this cone opening angle is small, a few degrees at most.

For a spinless particle, each plane wave component $|PW(\vec p)\rangle$ inside the vortex state is multiplied by the
phase factor $\exp(i \ell \varphi_p)$,
\begin{equation}
|\varkappa, \ell\rangle \propto \int d^2 \vec p_\perp\, \delta(|\vec p_\perp|-\varkappa) e^{i \ell \varphi_p}\, |PW(\vec p)\rangle
\propto \int d\varphi_p e^{i \ell \varphi_p}\, |PW(\vec p)\rangle\,,\label{vortex-spinless}
\end{equation}
which produces the phase vortex in the coordinate space: $\exp(i\ell\varphi_r)$ \cite{Jentschura:2010ap, Jentschura:2011ih}.
For particles with polarization degrees of freedom such as photons or fermions the construction is more subtle.
For fermions, one can construct exact monochromatic solutions of the Dirac equation \cite{Bliokh:2011fi, Karlovets:2012eu, Serbo:2015kia},
which are eigenstates of the total angular momentum $z$-projection operator $\hat j_z$
but not of spin $\hat s_z$ or OAM $\hat \ell_z$ operators individually.
The spin degrees of freedom in these solutions can be parametrized in a way similar to the vector $\vec s$ \cite{Bliokh:2011fi},
which, however, cannot be interpreted as the spin of vortex electron in the rest frame because there exists no reference
frame for all plane wave components simultaneously.
Alternatively, one can describe the polarization state of a vortex fermion with a given helicity which is
a conserved quantum number \cite{Karlovets:2012eu, Serbo:2015kia}.

Whatever the choice, one can define a twisted fermion in the same wave as in Eq.~\eqref{vortex-spinless}
bearing in mind that now each individual plane wave $|PW(\vec p)\rangle$ contains a bispinor $u_{p, s}$.
It must satisfy Dirac's equation and, therefore, it depends on the plane wave momentum $\vec p$ and cannot be taken out of the integral.

However, this dependence brings up not only technical difficulties but also novel opportunities
absent in the plane wave case.
Indeed, the spin degrees of freedom used to define $u_{p, s}$ can depend on $\vec p$ in a non-trivial way.
	As a result, a polarization state of a non-plane-wave fermion is, in general, described with a polarization {\em field} rather than polarization parameters.
For vortex fermions, the polarization field can exhibit polarization singularity along the phase singularity line.
Such singularities are well known for vector fields, in particular in optics, and were classied in \cite{Dennis:2002}.
For fermion fields, they were described in a similar way in \cite{Sarenac:2018} with applications to vortex neutrons.

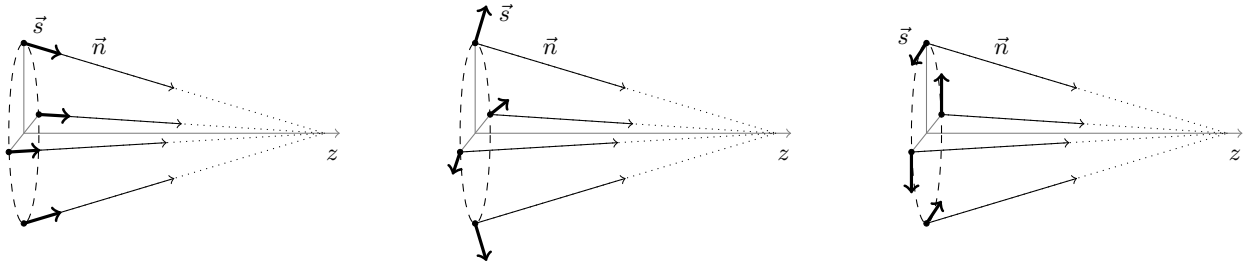
\begin{figure}[!h]
	\centering
	\begin{tikzpicture}
	\draw [->,gray,thin] (0,0) -- (4.2,0);
	\draw (4.1,-0.1) node[below] {$z$};
	\draw [gray,thin] (0,0) -- (0,1.2);
	\draw [gray,thin] (0.2,0.25) -- (-0.2,-0.25);
	\draw [dotted] (0,1.2) -- (4,0);
	\draw [dotted] (0,-1.2) -- (4,0);
	\draw [dotted] (-0.2,-0.25) -- (4,0);
	\draw [dotted] (0.2,0.25) -- (4,0);
	\draw[->] (0,1.2) -- (2,0.6);
	\draw[->] (0,-1.2) -- (2,-0.6);
	\draw[->] (-0.2,-0.25) -- (1.9,-0.125);
	\draw[->] (0.2,0.25) -- (2.1,0.125);

	\draw[->,very thick] (0,1.2) -- (0.5,1.05);
	\draw[->,very thick] (0,-1.2) -- (0.5,-1.05);
	\draw[->,very thick] (-0.2,-0.25) -- (0.22,-0.225);
	\draw[->,very thick] (0.2,0.25) -- (0.62,0.225);

	\draw [dashed] (0,0) ellipse (0.2 and 1.2);
	\filldraw [black] (0,1.2) circle (1pt);
	\filldraw [black] (0,-1.2) circle (1pt);
	\filldraw [black] (-0.2,-0.25) circle (1pt);
	\filldraw [black] (0.2,0.25) circle (1pt);

\draw (0.0,1.45) node[right] {$\vec s$};
\draw (1,1.4) node[below] {$\vec n$};

\draw [->,gray,thin] (6,0) -- (10.2,0);
\draw (10.1,-0.1) node[below] {$z$};
\draw [gray,thin] (6,0) -- (6,1.2);
\draw [gray,thin] (6.2,0.25) -- (5.8,-0.25);
\draw [dotted] (6,1.2) -- (10,0);
\draw [dotted] (6,-1.2) -- (10,0);
\draw [dotted] (5.8,-0.25) -- (10,0);
\draw [dotted] (6.2,0.25) -- (10,0);
\draw[->] (6,1.2) -- (8,0.6);
\draw[->] (6,-1.2) -- (8,-0.6);
\draw[->] (5.8,-0.25) -- (7.9,-0.125);
\draw[->] (6.2,0.25) -- (8.1,0.125);

\draw[->,very thick] (6,1.2) -- (6.15,1.7);
\draw[->,very thick] (6,-1.2) -- (6.15,-1.7);
\draw[->,very thick] (5.8,-0.25) -- (5.7,-0.55);
\draw[->,very thick] (6.2,0.25) -- (6.45,0.45);

\draw [dashed] (6,0) ellipse (0.2 and 1.2);
\filldraw [black] (6,1.2) circle (1pt);
\filldraw [black] (6,-1.2) circle (1pt);
\filldraw [black] (5.8,-0.25) circle (1pt);
\filldraw [black] (6.2,0.25) circle (1pt);

\draw (6.2,1.6) node[right] {$\vec s$};
\draw (7,1.4) node[below] {$\vec n$};

\draw [->,gray,thin] (12,0) -- (16.2,0);
\draw (16.1,-0.1) node[below] {$z$};
\draw [gray,thin] (12,0) -- (12,1.2);
\draw [gray,thin] (12.2,0.25) -- (11.8,-0.25);
\draw [dotted] (12,1.2) -- (16,0);
\draw [dotted] (12,-1.2) -- (16,0);
\draw [dotted] (11.8,-0.25) -- (16,0);
\draw [dotted] (12.2,0.25) -- (16,0);
\draw[->] (12,1.2) -- (14,0.6);
\draw[->] (12,-1.2) -- (14,-0.6);
\draw[->] (11.8,-0.25) -- (13.9,-0.125);
\draw[->] (12.2,0.25) -- (14.1,0.125);

\draw [dashed] (12,0) ellipse (0.2 and 1.2);
\filldraw [black] (12,1.2) circle (1pt);
\filldraw [black] (12,-1.2) circle (1pt);
\filldraw [black] (11.8,-0.25) circle (1pt);
\filldraw [black] (12.2,0.25) circle (1pt);

\draw[->,very thick] (12,1.2) -- (11.8,0.9);
\draw[->,very thick] (12,-1.2) -- (12.2,-0.9);
\draw[->,very thick] (11.8,-0.25) -- (11.8,-0.8);
\draw[->,very thick] (12.2,0.25) -- (12.2,0.8);

\draw (11.5,1.3) node[right] {$\vec s$};
\draw (13,1.4) node[below] {$\vec n$};
	\end{tikzpicture}
	\caption{Three examples of non-trivial polarization states for vortex fermions: parallel (left), radial (center), and azimuthal (right).
	Long thin arrows illustrate the momenta of individual plane wave components inside a vortex state; the thick short arrows
	indicate the vector $\vec s$ for each plane wave component.}
	\label{fig-polarization-states}
\end{figure}

In this exploratory study, we will consider three benchmark options for the vortex muon polarization states,
which are illustrated in Fig.~\ref{fig-polarization-states}.
\begin{itemize}
	\item
	{\bf Parallel polarization}. For each plane wave component with a given direction $\vec n$, we choose $\vec s = \vec n$.
	A motivation for this choice comes from the observation that,
	when muon moves in an external magnetic field, its spin precession approximately follows the momentum.
	Thus, if a vortex muon is obtained from a forward polarized plane wave muon through interaction with an external mangetic field,
	the parallel polarization state will result.
	\item
	{\bf Radial polarization}. For each plane wave component with a given $\vec n$, we consider the plane spanned by axis $z$ and $\vec n$
	and define the unit vector $\vec n_t$ via $\vec n_t \vec n = 0$, $(n_t)_z > 0$. The radial polarization corresponds to choosing $\vec s = \vec n_t$.
	\item
	{\bf Azimuthal polarization}: with $\vec n$ and $\vec n_t$ defined above, we construct the azimuthally pointing unit vector
	$\vec n_\varphi = \vec n \times \vec n_t$ and assume that $\vec s = \vec n_\varphi$.
\end{itemize}
We would like to make a few remarks concerning these states.
In the plane wave limit, the parallel polarization state approaches the muon with spin along axis $z$.
On the other hand, for the radial and azimuthal polarizations, there are no plane wave counterparts because
the definition of such states requires the presence of a polarization singularity line.
These states represent novel forms of polarization possible for non-plane-wave fermions.
In the plane wave limit they will disappear, corresponding to the unpolarized case.

When exploring these polarization states, we would like to stress that do not claim that all these states can be easily created.
We just use them to illustrate the sensitivity of the muon decay to the polarization state beyond plane waves.
Once vortex muons are created, one should perform a detailed study of the effects taking into account the realistic parameters
achieved in the experiment.

Finally, notice that all of these polarization states are azimuthally invariant.
As a consequence, the final electron angular distribution
will be azimuthally symmetric.
Of course, other polarization states of Bessel vortex fermions can be defined, see various examples
in \cite{Sarenac:2018}. In many of these cases, the azimuthal symmetry of the angular distribution is broken.
Although we will not analyze such states below, our formalism can be readily extended to any of them.

\subsection{Decay of unpolarized vortex muon: exact results}

When passing from the plane wave to vortex muon decay, we still use the plane wave basis for description
of the final state, not attempting to measure the possible vortex nature of the emitted electron
but just exploring, as before, its energy spectrum and angular distribution.

In this case we do not need to recalculate the decay process itself. We know that each plane wave component
inside the vortex muon decays to a final state with its own kinematic configuration and, as a result, different
plane wave components do not interfere. Thus, the decay width of a Bessel vortex muon can be written
as an azimuthal average of the corresponding plane wave decay widths \cite{Ivanov:2011kk}:
\begin{equation}
d{\Gamma} = \int \frac{d\varphi_p}{2\pi}\, d\Gamma_{PW}(\vec p)\,,\label{vortex-decay-1}
\end{equation}
where $d\Gamma_{PW}$ is given by \eqref{PWresult-3} or \eqref{PWresult-4}.
Notice that this decay rate becomes independent of the value of OAM. This is an unavoidable feature
in processes involving only one twisted particle in a single OAM state.
Dependence on the value of the OAM may arise only if we measure the final particle OAM 
(which represents an additional challenge in high-energy physics processes)
or if we prepare the initial state in a superposition of different OAM states.

However, even without the access to the value of the OAM, one can certainly explore its non-trivial
momentum distribution. It is the vortex cone structure which will lead to significant modifications
for the vortex muon with respect to the plane wave case.

Let us first consider the decay of an unpolarized vortex muon.
We choose the direction of the final electron
to define the $(x,z)$ plane: $\vec n_e = (\sin\theta,\, 0,\, \cos\theta)$.
In this coordinate system, any chosen plane wave component inside the vortex muon
is defined by $\vec n = (\sin\theta_0 \cos\varphi_p,\, \sin\theta_0\sin\varphi_p,\,\cos\theta_0)$,
where $\theta_0$ is the cone opening angle given in \eqref{cone}.
Then, at fixed $E_e$, the angular dependence comes from the factor
\begin{equation}
1-\beta\, \vec n\vec n_e = a - b \cos\varphi_p\,, \quad \mbox{where}\quad 
a = 1 - \beta \cos\theta\cos\theta_0\,, \quad b = \beta \sin\theta\sin\theta_0 \,.\label{ab}
\end{equation}
This factor changes between $1 - \beta \cos(\theta-\theta_0)$ to $1 - \beta \cos(\theta+\theta_0)$.
According to Eq.~\eqref{Emax}, for any electron angle $\theta_0$,
these plane wave components produce {\em different} maximal energies of the electron, spanning the interval
\begin{equation}
\mbox{from}\quad E_{e1} = \frac{m^2}{2E (1-\beta\cos(\theta+\theta_0))}\quad \mbox{to} \quad
E_{e2} = \frac{m^2}{2E (1-\beta\cos(\theta-\theta_0))}\,.\label{Emax12}
\end{equation}
As a result, the electron spectrum acquires additional features.
For any given $\theta$, it contains two regions which differ by the range of plane wave components
inside the vortex muon contributing to the decay:
\begin{eqnarray}
\mbox{lower energy region:}& E_e < E_{e1}\,,& \mbox{contributing $\varphi_p$:} \quad 0 \le \varphi_p \le 2\pi\,,\nonumber\\
\mbox{higher energy region:}& \quad E_{e1} \le E_e \le E_{e2}\,, \quad & \mbox{contributing $\varphi_p$:} \quad -\tau \le \varphi_p \le \tau\,.\label{two-regions}
\end{eqnarray}
Here, the azimuthal range $\tau$ is given by the inverse of Eq.~\eqref{Emax} and can be presented as
\begin{equation}
\cos\tau = \frac{(E_e-E_{e1})E_{e2} - (E_{e2}-E_e)E_{e1}}{E_e(E_{e2}-E_{e1})}\,. \label{tau}
\end{equation}
One can see that, as $E_e$ goes from $E_{e1}$ to $E_{e2}$, the value of $\cos\tau$ goes from $-1$ to 1.

If $\theta > \theta_0$ (detecting electrons outside of the vortex cone),
the spectrum extends to {\em higher} energies than for the plane wave muon of the same energy.
Indeed, one sees that the end-point of the plane wave spectrum $E_{e\,\rm max}$ would lie between $E_{e1}$ and $E_{e2}$.
Angles $\theta < \theta_0$, corresponding to the region inside the cone, which is unavailable for the plane wave case,
display a more intricate situation. The upper energy limit $E_{e2}$ can be either higher than $E_{e\,\rm max}$ (for $\theta > \theta_0/2$)
or lower than $E_{e\,\rm max}$ (for $\theta < \theta_0/2$). Thus, the spectral-angular distribution is predicted to have a forward dip
for the vortex muon, in sharp contrast with the plane-wave muons.

Next, we define the spectral-angular distribution $w(E_e, \theta, \varphi)$ and the angular distribution $W(\theta, \varphi)$ in the following way:
\begin{equation}
\frac{d\Gamma}{dE_ed\Omega}= \frac{\Gamma_0}{\gamma}\, w(E_e, \theta, \varphi)\,, \quad 
\frac{d\Gamma}{d\Omega} = \frac{\Gamma_0}{\gamma}\, \int w(E_e, \theta, \varphi) dE_e = \frac{\Gamma_0}{\gamma}\, W(\theta, \varphi)\,,
\quad \int W(\theta, \varphi) d\Omega = 1\,.
\label{distributions-defined}
\end{equation}
The angular distribution $W(\theta,\varphi)$ is dimensionless, while the spectral-angular distribution $w(E_e, \theta, \varphi)$, 
or simply the spectrum if only the fixed-angle energy distribution is concerned, has dimension GeV$^{-1}$. 
Since the distributions will always be azimuthally symmetric in the examples we consider in this paper,
we will suppress $\varphi$ to simplify the notation and write $w(E_e, \theta)$ and $W(\theta)$.
However, the normalization conditions \eqref{distributions-defined} are always assumed.

Now, substituting the unpolarized version of the plane-wave decay rate \eqref{PWresult-3} in \eqref{vortex-decay-1} 
and performing the $\varphi_p$ integration within the limits \eqref{two-regions}, we obtain the 
following generic expression for the spectral-angular distribution:
	\begin{equation}
w(E_e, \theta) = \frac{4\gamma E_e^2 }{\pi m^5} \, (3m^2 \cdot A - 4 E E_e \cdot B)\,,
\end{equation}
with the coefficients $A$ and $B$ displaying different behavior in the two energy regions.
In the lower energy region, we get
\begin{equation}
A = a\,, \quad B = a^2 + \frac{1}{2}b^2\,,
\end{equation}
with angular variables $a$ and $b$ defined in \eqref{ab}. These coefficients do not depend on the electron energy. 
For the higher energy region, the range of the $\varphi_p$ integration depends on $E_e$ through
$\tau(E_e)$ given by \eqref{tau}, and we get
\begin{equation}
A(E_e) = a \frac{\tau}{\pi} - b\frac{\sin\tau}{\pi}\,, \quad
B(E_e)= \left(a^2 + \frac{1}{2}b^2\right)\frac{\tau}{\pi} - 2ab\, \frac{\sin\tau}{\pi} + b^2\frac{\sin2\tau}{4\pi}\,.
\end{equation}
If we do not measure the energy of the electron and want just to explore its angular distribution,
we can integrate over all electron energies available for a fixed $\theta$.
For this calculation, it is convenient to insert the unpolarized version of Eq.~\eqref{PWresult-4} into Eq.~\eqref{vortex-decay-1} and
perform the $\varphi_p$ integral.
Here and below, we will make use of the integrals
\begin{eqnarray}
{\cal I}_2 &=& \frac{1}{2\pi}\int d\varphi_p \frac{1}{(a - b\cos\varphi_p)^2} = \frac{a}{(a^2-b^2)^{3/2}}\,,\nonumber\\[1mm]
{\cal I}_3 &=& \frac{1}{2\pi}\int d\varphi_p \frac{1}{(a - b\cos\varphi_p)^3} = \frac{1}{2}\frac{2a^2+b^2}{(a^2-b^2)^{5/2}}\,, \label{integrals}
\end{eqnarray}
with the same $a$ and $b$ as in \eqref{ab}.
With these intergrals, we get
\begin{equation}
W(\theta) = 
\frac{1}{4\pi \gamma^2}\, {\cal I}_2 =
\frac{1}{4\pi\gamma^2}\,\frac {1 - \beta \cos\theta\cos\theta_0}{\bigl\{ [1 - \beta \cos(\theta+\theta_0)][1 - \beta \cos(\theta-\theta_0)]\bigr\}^{3/2}}\,.\label{vortex-result-2} 
\end{equation}
Notice the invariance of this result under the exchange $\theta_0 \leftrightarrow \theta$.

In the non-relativistic limit $\beta \ll 1$, we recover the usual isotropic decay rate.
Notice that the exact limit $\beta \to 0$ cannot be taken for vortex muons due to the non-zero $\varkappa$.
Even if one performs a Lorentz boost and sets $p_z=0$, one gets a non-zero value $\beta = \varkappa/\sqrt{\varkappa^2+m^2}$.
In the opposite limit of the ultrarelativistic muon $\beta \rightarrow 1$, $\gamma \gg 1$, we get
a narrow angular distribution peaked at $\theta = \theta_0$. The exact integration
over all angles still yields $\Gamma = \Gamma_0/\gamma$.
Finally, in the plane wave limit $\theta_0 \to 0$, we recover Eq.~\eqref{PWresult-4}.

\subsection{Decay of unpolarized vortex muon: numerical examples}

To illustrate the above analysis and to stress the modifications expected for vortex muons,
let us fix the muon energy at $E = 3.1$ GeV, which corresponds to the ``magic value'' $\gamma = 29.3$
used in all modern $g-2$ experiments \cite{Miller:2007kk, Logashenko:2018pcl}.
For the cone opening angle $\theta_0$, we choose two benchmark values:
\begin{equation}
\mbox{narrow cone:}\quad \theta_0 = 0.01 \approx 0.6^\circ\,, \qquad
\mbox{wide cone:} \quad \theta_0 = 0.1 \approx 6^\circ.
\label{two-cones}
\end{equation}
We consider the narrow cone to be closer to the realistic values, but we will also show results
for the wide cone to highlight strong distortions of the distributions.

\begin{figure}[!h]
	\centering
	\includegraphics[width=0.48\textwidth]{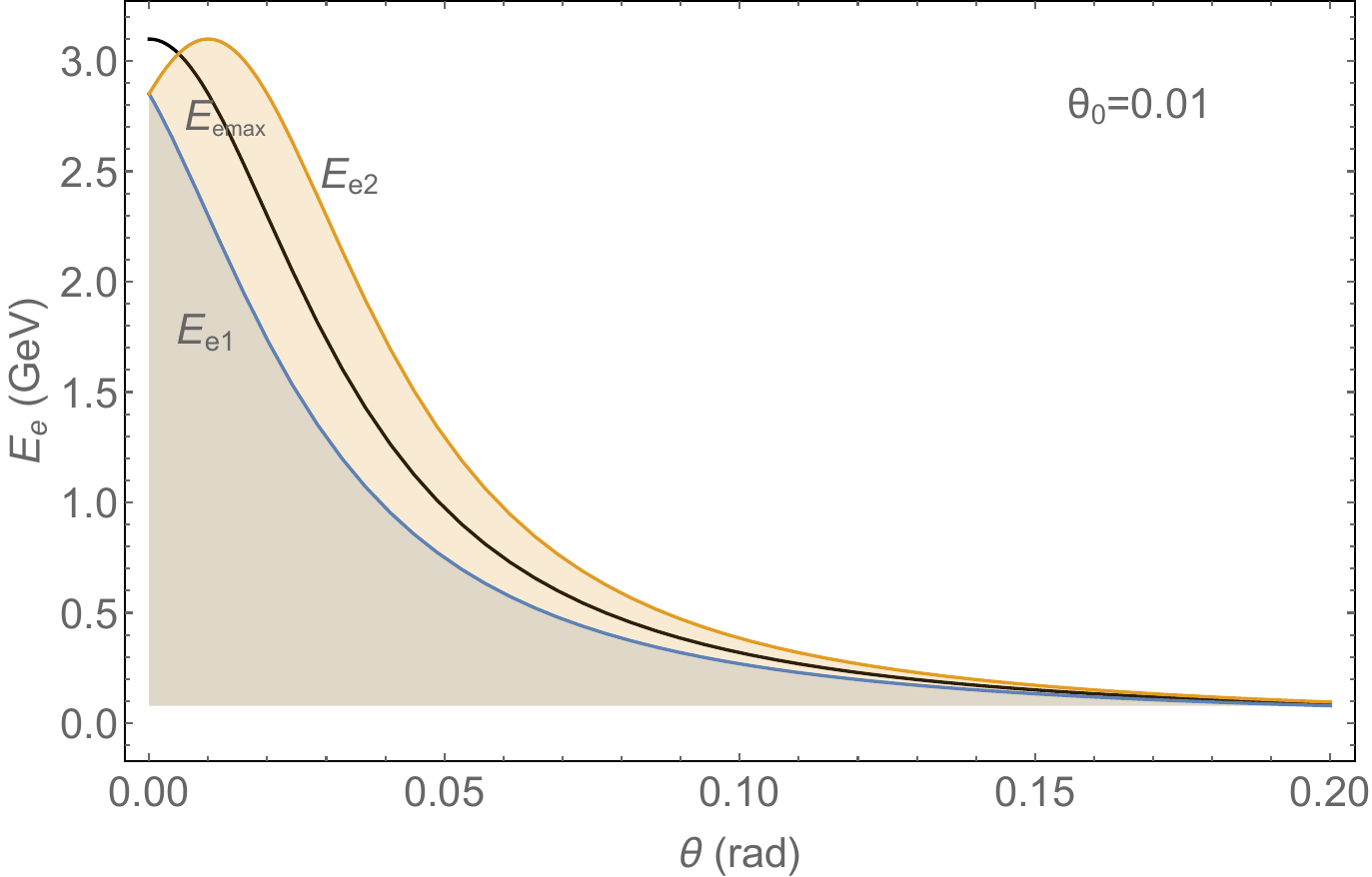}\hfill
	\includegraphics[width=0.48\textwidth]{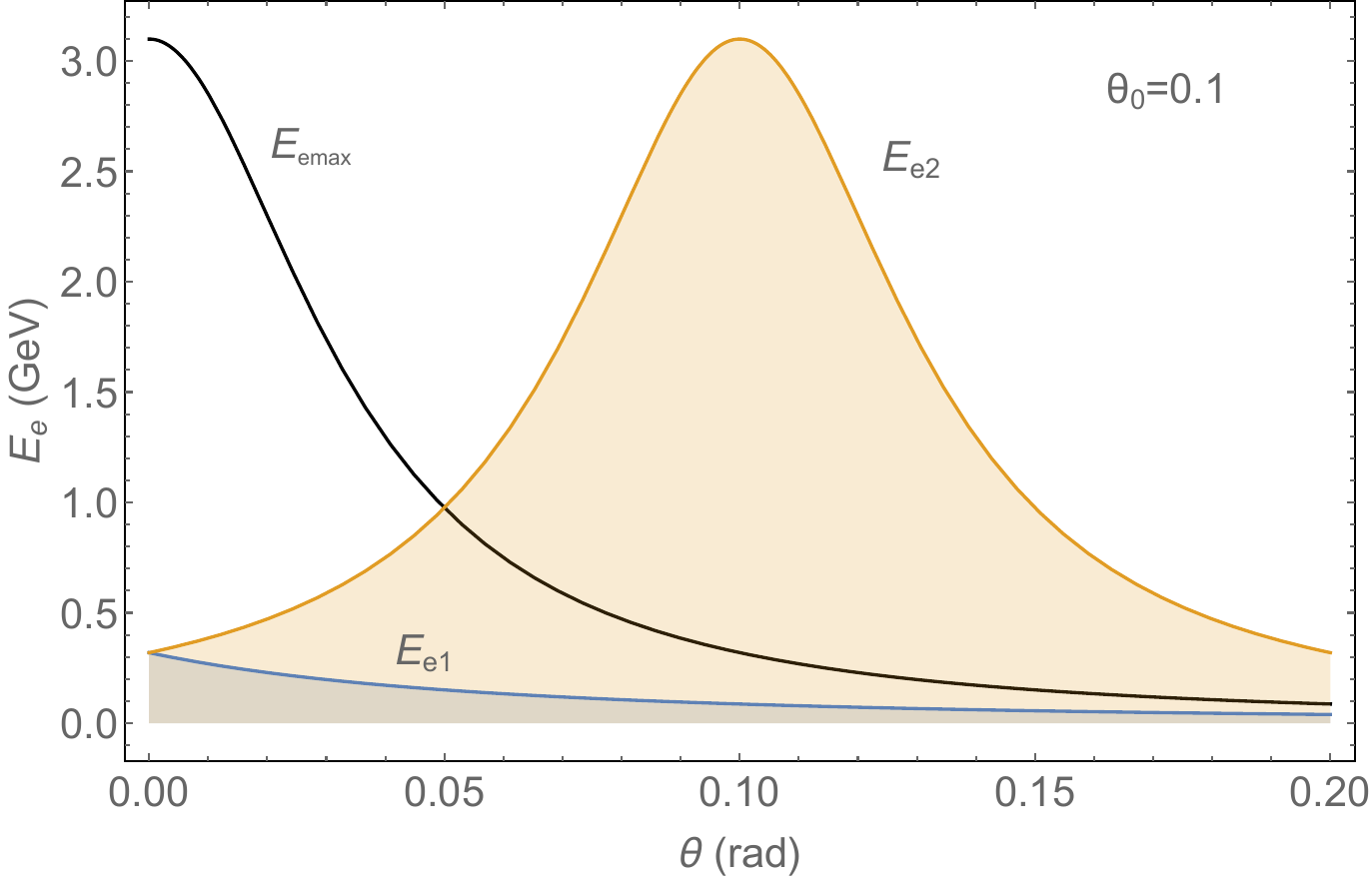}
	\caption{The electron energy range for the decay of 3.1 GeV muons as a funuction of the electron emission angle $\theta$ in the forward region.
		The black line corresponds to $E_{e\,\rm max}$ from the plane wave muon.
		The two colored lines correspond to $E_{e1}$ and $E_{e2}$ from the vortex muon with
		$\theta_0 = 0.01$ (left) and 0.1 (right).}\label{Fig-Emax}
\end{figure}

Since for ultrarelativistic muons, the angular distribution of the emitted electron develops a sharp forward peak,
from now on we focus on the forward region $\theta < 0.2 \approx 12^\circ$.
In Fig.~\ref{Fig-Emax} we plot the angular dependence of the energies $E_{e1}$ and $E_{e2}$ defined in \eqref{Emax12}
as functions of the electron emission angle $\theta$ in the forward region and compare them with the maximal electron energy
for the plane wave muon case \eqref{Emax}.
The left and right plots correspond to the narrow and wide opening angles \eqref{two-cones}.

The key observation here is that, even for the narrow cone with $\theta_0 \ll 1$,
we observe a very significant shift of the two energies with respect to the plane wave case.
The ``higher energy'' region of the spectrum, shown in these plots with a lighter shading,
represent a sizable part of the total energy range.
In order to see what part of the spectrum is occupied by this new region, we compute
\begin{equation}
\frac{E_{e2}-E_{e1}}{E_{e2}+E_{e1}} = \frac{\beta\sin\theta\sin\theta_0}{1-\beta\cos\theta\cos\theta_0} \approx 2\frac{\theta_0}{\theta}\,,
\end{equation}
where the last estimate assumes $\beta \to 1$.
We conclude that even for very narrow vortex cones, we have access to a parametrically large part of the
electron spectrum, provided we can measure electrons emitted at angles $\theta$ comparable to $\theta_0$.
This part of the spectrum is unavailable in the plane wave case and represents a new kinematic feature of the vortex muon.

For wider cones, the difference between $E_{e1}$, $E_{e2}$, and the plane wave end point $E_{e\,\rm max}$ is dramatic, Fig.~\ref{Fig-Emax}, right.
For $\theta \sim \theta_0$, the energy spectrum extends far beyond what one would get, at these angles, from plane wave muons.
This is, of course, no surprise since the vortex muon contains plane wave components pointing close to the observation direction.

\begin{figure}[!h]
	\centering
	\includegraphics[width=0.48\textwidth]{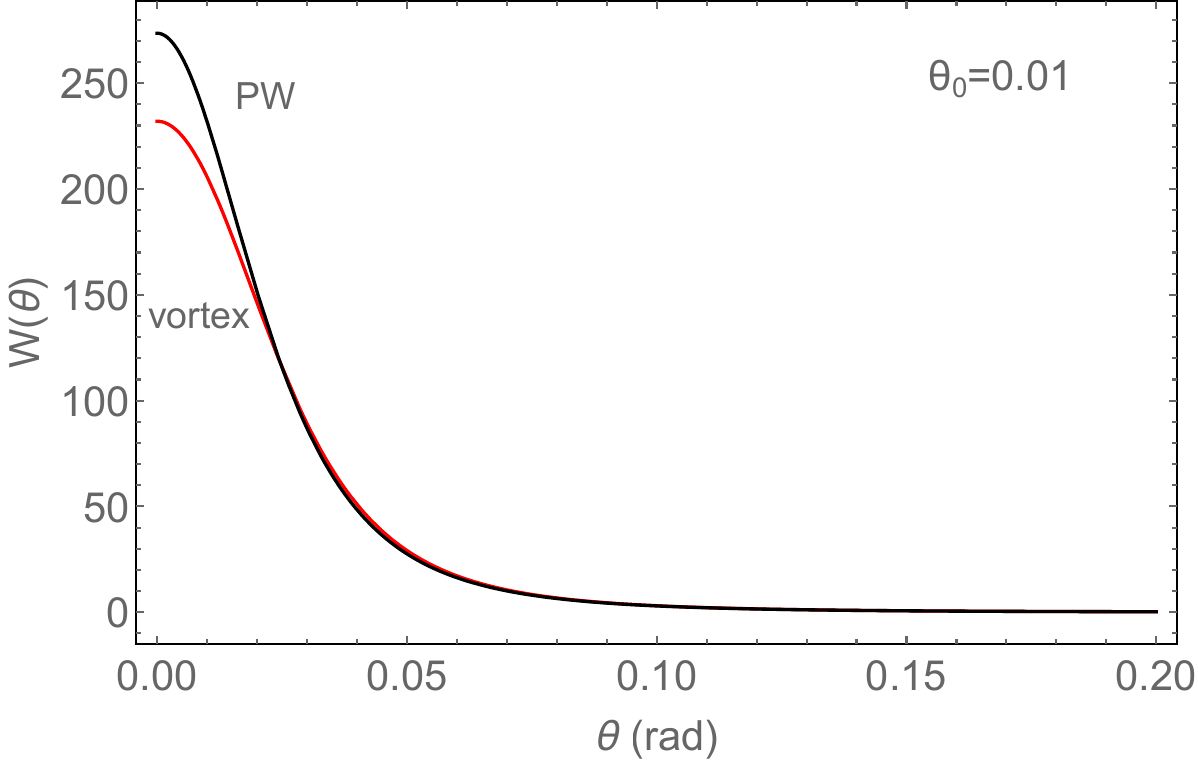}\hfill
	\includegraphics[width=0.48\textwidth]{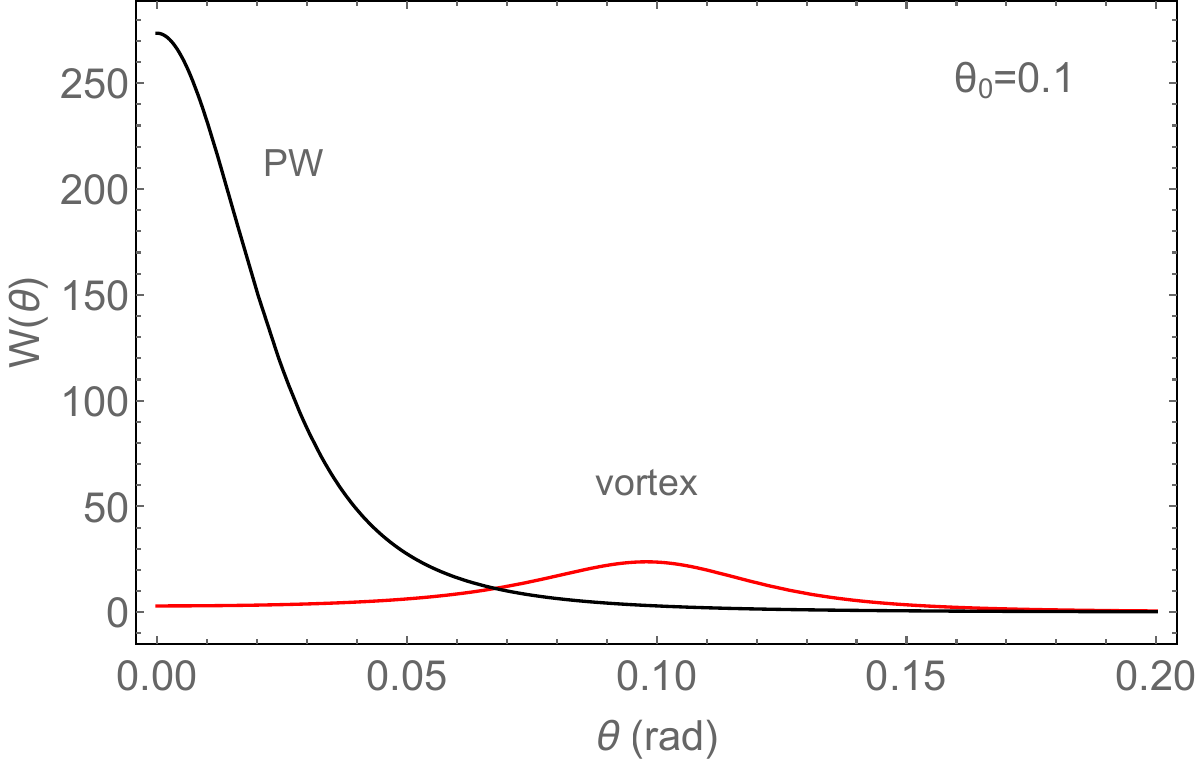}
	\caption{Electron angular distributions $W(\theta)$ emitted from the $3.1$ GeV vortex muon decays near the forward direction
	for $\theta_0 = 0.01$ (left) and 0.1 (right), plotted in red, compared with the plane wave muons (black lines).}\label{Fig-angular}
\end{figure}

Next we check how the electron angular distribution, integrated over energies,
differs with respect to the plane wave case.
In Fig.~\ref{Fig-angular} we plot the angular distributions \eqref{vortex-result-2}
in the forward region for the same two values of the cone opening angle.
For the realistic narrow cone, we just observe minor angular broadening,
which could also be mimicked by energy and angular smearing.
Thus, if we want to distinguish the vortex muon with a small opening angle from the plane wave muon,
the angular distribution alone is not a particularly revealing observable.
For a wide cone, the picture changes: we see a pronounced forward dip, as the angular distribution
peaks at $\theta = \theta_0$ and gets significantly depleted near $\theta = 0$.

\begin{figure}[!h]
	\centering
	\includegraphics[width=0.48\textwidth]{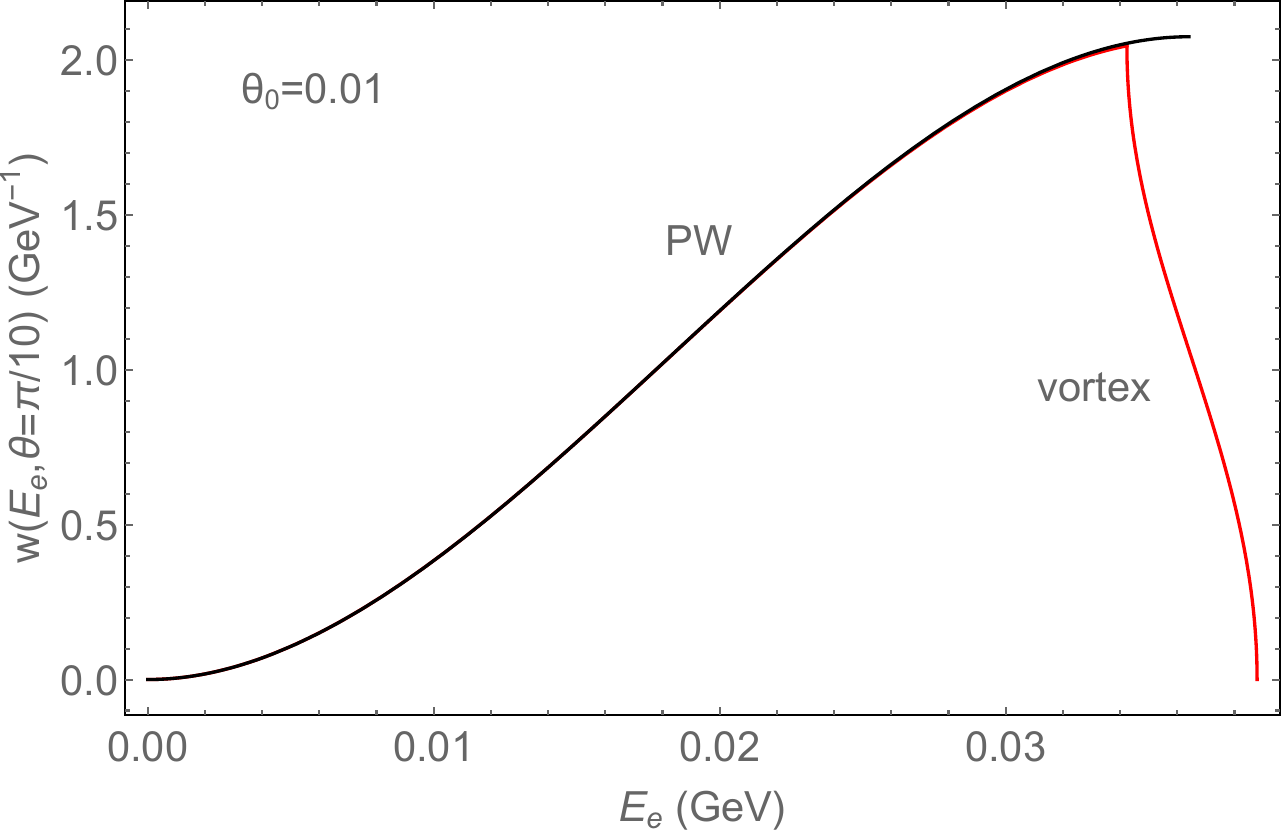}\hfill
	\includegraphics[width=0.48\textwidth]{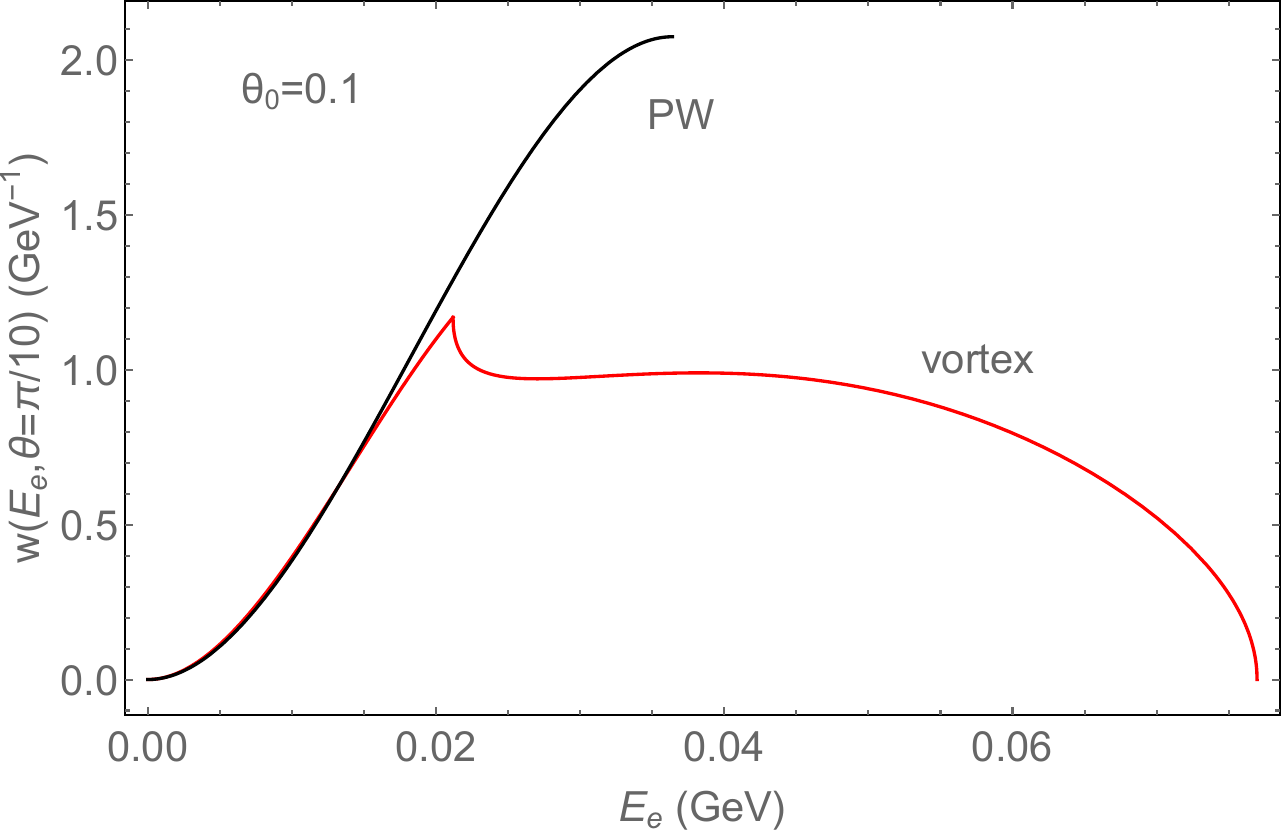}
	\caption{Electron energy spectrum from $3.1$ GeV vortex muon decays, $w(E_e, \theta)$ calculated at fixed $\theta = \pi/10 = 18^\circ$,
		for $\theta_0 = 0.01$ (left) and 0.1 (right) compared with the plane wave muons.}\label{Fig-energy}
\end{figure}

Since the angular distribution alone is not particularly useful, we turn to the energy spectrum
of electrons $w(E_e, \theta)$ detected at a given emission angle $\theta$.
In Fig.~\ref{Fig-energy} we show how the electron spectrum is modified with respect to the plane wave muons.
The two plots correspond to the same electron polar angle $\theta = \pi/10 = 18^\circ$ with the two choices
for the opening cones \eqref{two-cones}.

Even for the narrow vortex cone, the modifications are quite substantial, Fig.~\ref{Fig-energy}, left.
The plane wave muon would result in a smoothly rising spectrum extending to $E_{e\, \rm max} = 36$ MeV for this observation angle.
In the vortex muon case with $\theta_0 = 0.01$, the spectrum follows this behavior in the lower energy region.
However, in the higher energy region, the spectrum exhibits a break and steadily drops to zero.
Notice that this higher energy region is quite large, extending from 34 to 39 MeV,
representing more than 10\% of the total energy range.

We conclude that the key observable is the energy spectrum at a small fixed angle, not the angular distribution.
It is this spectrum in its higher energy range that displays the clearest
distinction between the (approximately) plane wave and the vortex muons.

For larger opening cone values, we see a dramatic distortion of the spectrum, Fig.~\ref{Fig-energy}, right.
Instead of smoothly rising to the end-point, the spectrum first develops a cusp at $E_{e1} = 20$ MeV,
then developes a broad plateau, and finally drops to zero at the new end-point $E_{e2} = 77$ MeV.

\begin{figure}[!h]
	\centering
	\includegraphics[width=0.48\textwidth]{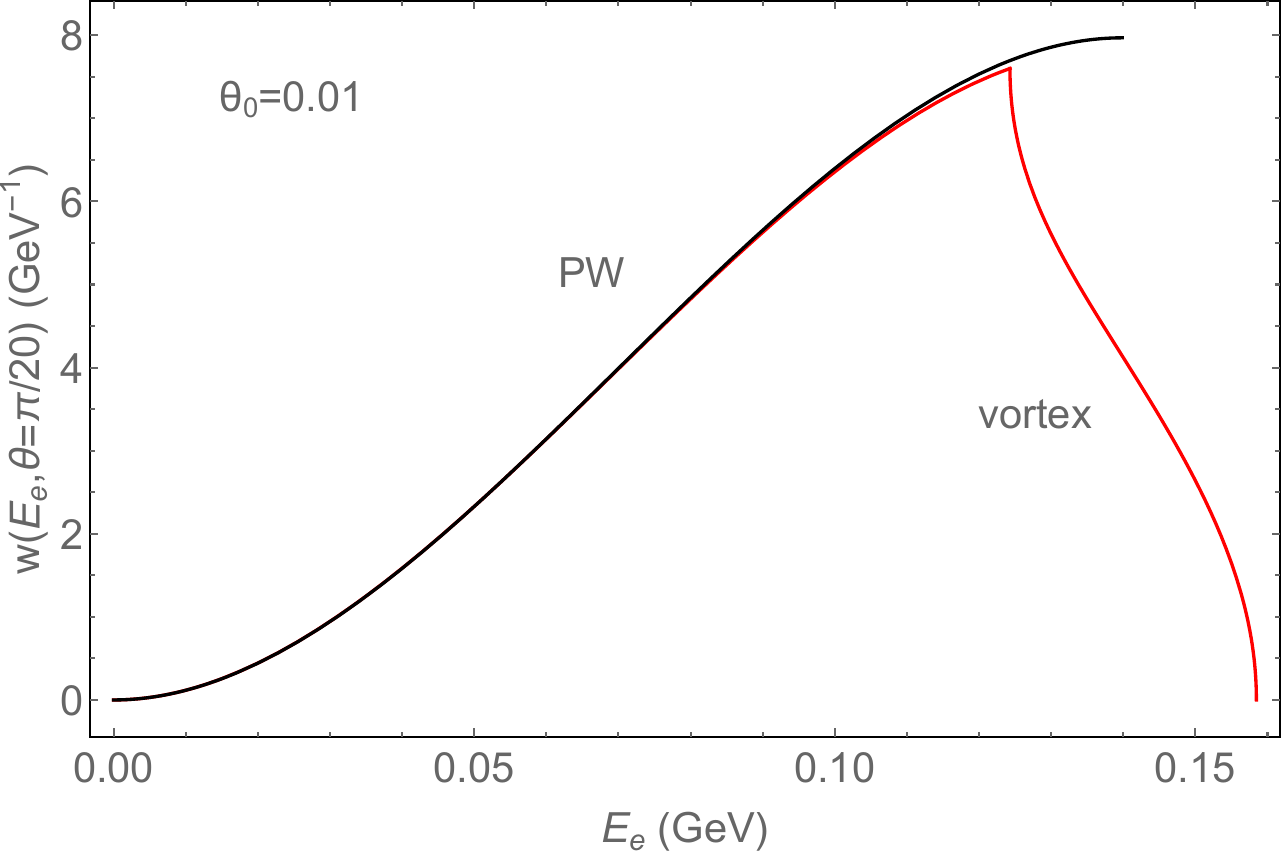}\hfill
	\includegraphics[width=0.48\textwidth]{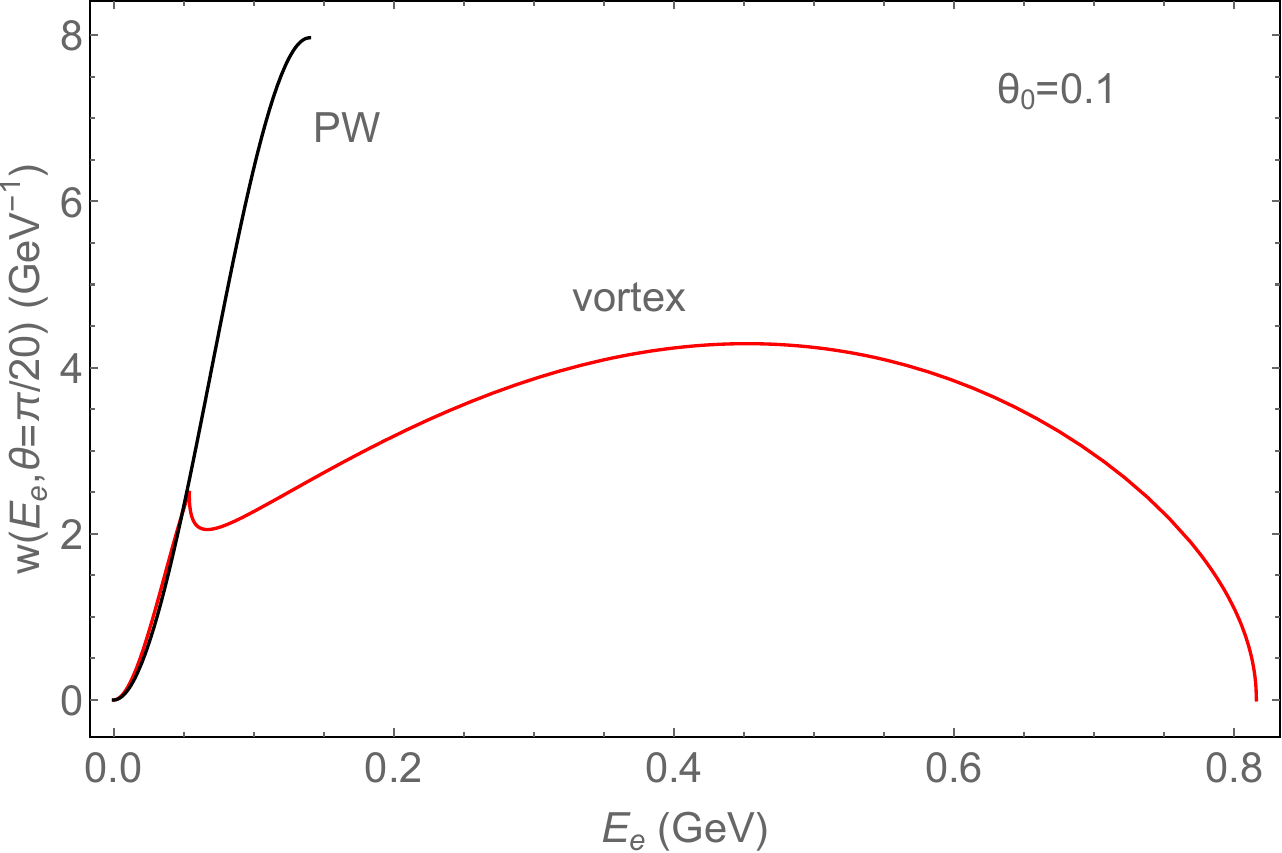}
	\caption{The same as in Fig.~\ref{Fig-energy} but for $\theta = \pi/20 = 9^\circ$.}\label{Fig-energy2}
\end{figure}

For smaller observation angles $\theta$, the difference becomes more pronounced. In Fig.~\ref{Fig-energy2} we plot the same energy spectra
but for $\theta = \pi/20 = 9^\circ$. Detecting such a behavior would be an unmistakable indication of the cone structure of the muon wave function.

\subsection{Decay of polarized vortex muon}

The decay rate of a polarized vortex muon is given by the same Eq.~\eqref{vortex-decay-1} with
$d\Gamma_{PW}$ taken from \eqref{PWresult-3} for the spectral-angular distribution of the electron $w(E_e,\theta)$
or from \eqref{PWresult-4} if the energy-integrated angular distribution $W(\theta)$ is concerned.
To specify the polarization state of the vortex muon, we need to
explicitly give the dependence of $\vec s$ on the plane wave momentum $\vec p$.
Let us begin with the energy-integrated angular distribution
and go through the polarization states outlined listed in Section~\ref{subsection-describing}.
\begin{itemize}
	\item
	The parallel polarization implies $\vec s = \vec n$, so that the plane wave decay rate \eqref{PWresult-4} now simplifies to
	\begin{equation}
	d\Gamma_{PW} = \Gamma_0\, \frac{d\Omega}{4\pi} \, \frac{1}{\gamma^3}\, \frac{1}{(1-\beta\, \vec n\vec n_e)^2}
	\left[1 + \frac{\beta - \vec n \vec n_e}{3(1-\beta\, \vec n\vec n_e)}\right]\,.
	\label{PWresult-5.1}
	\end{equation}
	As a result, the angular distribution has the following form
	\begin{equation}
	W(\theta) = \frac{1}{4\pi \gamma^2}\, \left[{\cal I}_2 \left(1 + \frac{1}{3\beta}\right)
	- {\cal I}_3 \frac{1}{3\beta\gamma^2}\right]\,,\label{vortex-result-3}
	\end{equation}
	where the integrals ${\cal I}_2$ and ${\cal I}_3$ are given in Eqs.~\eqref{integrals}.
	In the plane-wave limit, one must recover the plane wave result \eqref{PWresult-4.1}.
	\item
	For the radial and azimuthal polarization states, one has $\vec s \vec n = 0$ for each plane wave component.
	In this case, the plane wave decay rate simplifies to
	\begin{equation}
	d\Gamma_{PW} = \Gamma_0\, \frac{d\Omega}{4\pi} \, \frac{1}{\gamma^3}\, \frac{1}{(1-\beta\, \vec n\vec n_e)^2}
	\left[1 - \frac{\vec s \vec n_e}{3\gamma (1-\beta\, \vec n\vec n_e)}\right]\,.
	\label{PWresult-5.2}
	\end{equation}
	After integration over all the plane wave components of the vortex muon,
	we get the angular distribution for the radial polarization state
	\begin{equation}
	W(\theta) = \frac{1}{4\pi \gamma^2}\, \left[{\cal I}_2 \left(1 - \frac{\cos\theta_0}{3\beta\gamma \sin\theta_0}\right)
	+ {\cal I}_3 \frac{\cos\theta_0 - \beta \cos\theta}{3\gamma\beta\sin\theta_0}\right]\,.\label{vortex-result-4}
	\end{equation}
	For the azimuthal polarization state we get $\vec s \vec n_e = \sin\theta \sin \varphi_p$.
	Since the other factors in the integrand depend on $\cos\varphi_p$, we see that the spin contribution
	of the azimuthally polarized vortex muon vanishes after the $\varphi_p$ integration.
	Thus, the azimuthally polarized vortex muon will produce the angular distribution just as in the unpolarized case.
\end{itemize}

\begin{figure}[!h]
	\centering
	\includegraphics[width=0.55\textwidth]{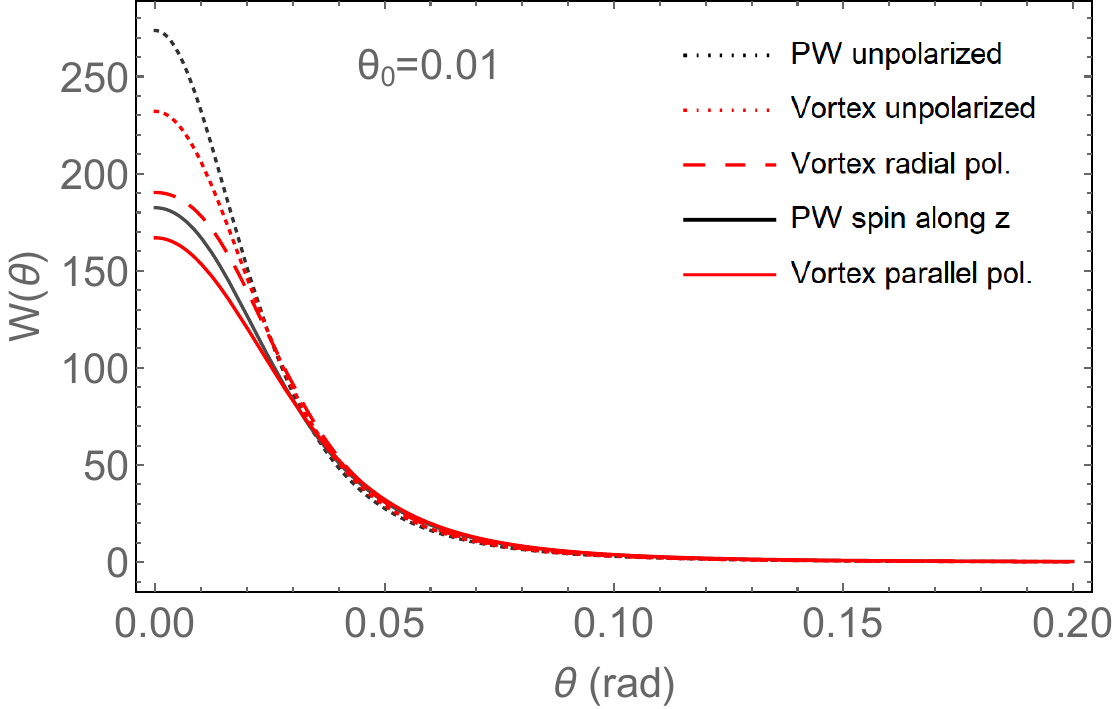}
	\caption{The electron angular distribution $W(\theta)$ emitted from plane wave (black) and vortex (red) muons in various polarization cases:
		unpolarized (dotted lines), parallel polarization (solid lines), radial polarization (dashed line).
		The cone opening angle is $\theta_0 = 0.01$.}\label{Fig-angular-polarized}
\end{figure}

In Fig.~\ref{Fig-angular-polarized} we plot the above angular distributions in the forward region for the narrow cone with $\theta_0 = 0.01$.
Here, the plane wave muon cases are shown in black, vortex muon cases are shown in red.
The dotted lines, both for plane wave and vortex muons, correspond to the unpolarized case;
they are identical to Fig.~\ref{Fig-angular}, left.
The solid lines correspond to the parallel polarization for vortex muon and the $z$ polarization for the plane wave muon.
The red dashed line shows the radial polarization case, which is possible only for vortex muon.
We observe here the same pattern as before: when passing from plane wave to vortex muons with very small cone opening angle $\theta_0$,
we just see the forward peak to get slightly broader.
Thus, the main message from this plot is that the spectrum-integrated angular distribution is not of much help when distinguishing different polarization states.

\begin{figure}[!h]
	\centering
	\includegraphics[width=0.55\textwidth]{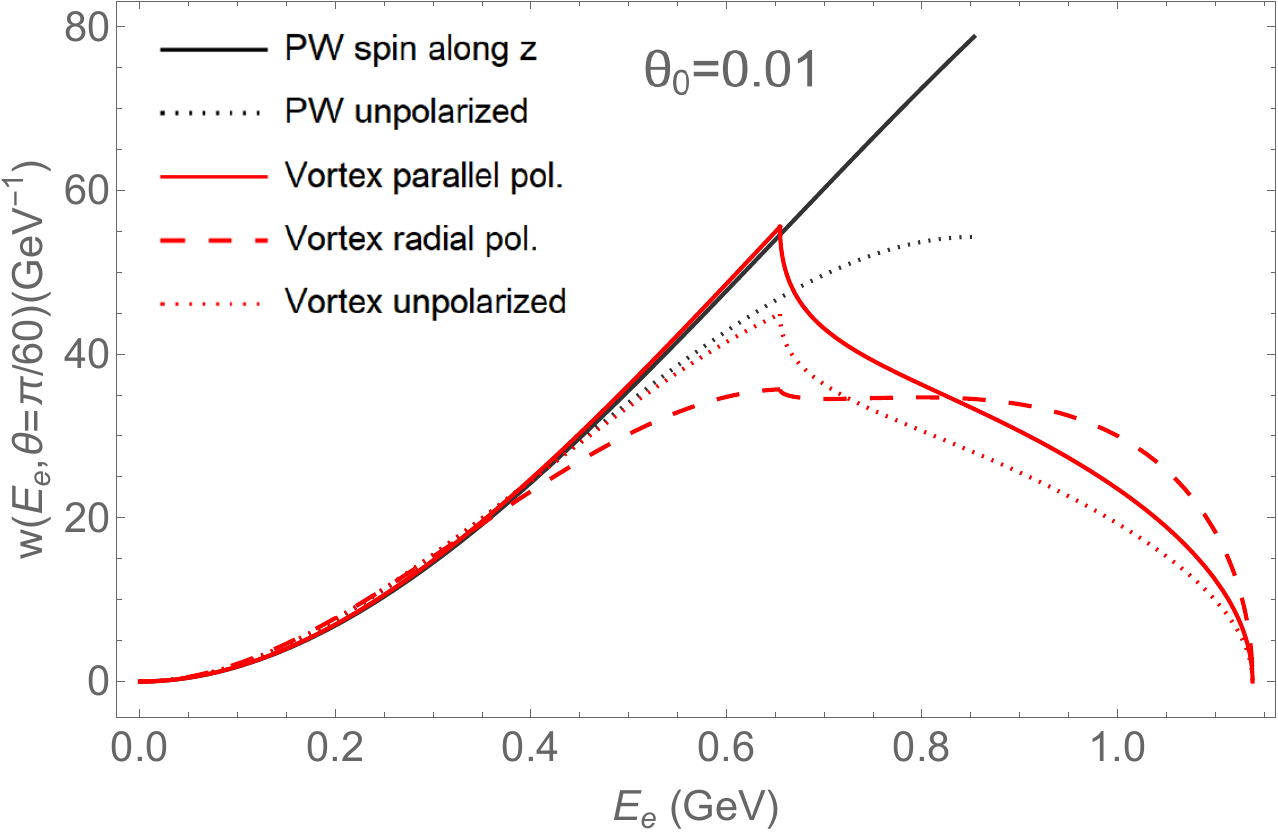}
	\caption{The electron spectra $w(E_e, \theta)$ at the observation angle $\theta = \pi/60 = 3^\circ$
		for plane wave (black) and vortex (red) muons in various polarization cases:
		unpolarized (dotted lines), parallel polarization (solid lines), radial polarization (dashed line).
		The cone opening angle is $\theta_0 = 0.01$.}\label{Fig-spectra-polarized}
\end{figure}

The distinction becomes much more clear when we look into the energy spectrum, especially in the higher energy region.
In Fig.~\ref{Fig-spectra-polarized} we show these spectra\footnote{We are grateful to Rahul Singh 
	and his colleagues for calling our attention to the erroneous spectra in an earlier version of this plot.} for the same five states:
unpolarized plane wave and vortex muons, parallel polarization for plane wave and vortex muons,
and the radial polarization. The spectra are given for the electron emission angle of $\pi/60 = 3^\circ$.
We see critically different patterns for various polarization options in the higher energy region,
which occupies at this observation angle the region from about $E_{e1}=700$ to $E_{e2}=1200$ MeV.

In particular, the spectrum of the electrons from radially polarized muons, which closely follows the unpolarized case below $E_{e1}$,
displays a strong second peak at high energies. The explanation of this effect is straightforward.
For lower energies, all plane wave components of the radially polarized muon contribute to spectrum,
and since all of them point in different directions, the end result looks almost like the unpolarized case.
However, in the higher energy region, only some of the plane wave components contribute, see Eq.~\eqref{two-regions}.
Therefore, the closer to the end point $E_{e2}$, the more the process resembles the transversely polarized case.

\section{Discussion and conclusions}\label{section-discussion}

Exploring collisions and decays of elementary particles prepared in vortex states with a non-zero orbital angular momentum
is a novel promising way to probe particle structure and dynamics.
Kinematic dependences, angular distributions, and spin effects have been predicted to
differ in a significant way with respect to the (approximate) plane wave case.
These predictions still await experimental verification, mostly due to the absence of suitable instrumentation
which would enable such studies. In order to stimulate instrumentation development,
one can theoretically explore what new features could become observable in experiments with vortex particles.

In this paper, we studied, for the first time, the peculiar features
which could be seen in decays of muons prepared in a vortex state.
We obtained the spectral-angular distribution of the emitted electron, both for unpolarized and polarized muons.
In the latter case, we took into account the richer list of opportunities which exists
for non-plane-wave fermions.

The hallmark feature of vortex particles is their ``cone structure'' in momentum space.
Thus, almost all studies of vortex particle collisions look into the modifications of the angular distribution
of the final state. However, when the cone opening angle $\theta_0$ is small,
these modifications with respect to the plane wave case are usually minor.

The key finding of this work is that it is the higher energy part of the electron spectrum, not the angular distribution,
that displays the strongest differences with respect to the plane wave muons and that
allows one to distinguish various polarization states.
Even for very small opening angles used in our numerical examples
we could observe very significant modifications of the energy spectrum, if the electron observation
angle $\theta$ is also small.
Fig.~\ref{Fig-spectra-polarized} offers a clear illustration:
for muons with $E= 3.1$ GeV and $\theta_0 = 0.01 \approx 0.6^\circ$ and for the electron observation angle $\theta = 3^\circ$,
we predict electron spectra to dramatically differ
for plane wave and vortex muons, as well as for different vortex muon polarization states.
In a sense, these findings confirm once again that muon decay is a self-analyzing process,
as it offers a clear information on the muon polarization state even beyond the plane wave case.

In this paper, we left aside the issue of producing vortex muons.
No experiment has tried it so far. However, we see no fundamental obstacles to producing such states,
although we admit that it requires instrumentation which is not presently available.
It is true that traditional schemes for generating vortex states based on fork diffraction gratings
are not suitable for muons due to their large penetration depth. However, muons are charged particles
and, therefore, they can be manipulated with electric and magnetic fields.
If a nearly plane-wave muon passes through an aperture with an artificial magnetic monopole
(the tip of a magnetized needle), it will acquire orbital angular momentum proportional to the effective magnetic charge,
which has already been demonstrated for electrons \cite{Beche:2014}.
Alternatively, if muons can be emitted from a pointlike source placed inside a wide solenoid,
then, at the exit of solenoid, the muon wave function can acquire an OAM, just as it was predicted 
for electrons \cite{Floettmann:2020uhc,Karlovets:2020tlg}.
Thus, ``twisting'' muons is technically possible, provided initial nearly plane wave muons can be created
with a sufficient transverse coherence length.

Once vortex muons are produced and accelerated, one could use the standard electron spectrometers and detectors
to explore the subtle features of vortex muons.
As demonstrated in this paper, the small electron observation angle $\theta$ serves as a magnification tool
for discerning the features which exist near the narrow cone of the vortex muon state.

Once a proof-of-principle experiment detects these modifications, a research program can begin
which would explore spin and OAM evolution of vortex muons in external magnetic fields.
There exist theoretical studies of how vortex electron could behave in external fields, in particular, 
in the transverse magnetic field of a storage ring \cite{Gallatin:2012ai}.
This evolution may be difficult to track for electrons.
However, vortex muons, thanks to the self-analyzing nature of their decays, may offer a better glimpse
on this evolution.

Finally, we mention that, since the main effects discussed here are kinematic, 
we expect that similar easily observable features may appear
in decays of other unstable relativistic particles.

\section*{Acknowledgments}

We thank the referee for careful reading and suggestions.
This work was supported by grants of the National Natural Science Foundation of China 
(Grant No. 11975320) and the Fundamental Research Funds for the Central Universities, Sun Yat-sen University.

\end{document}